\documentclass[prb,aps,reprint,citeautoscript,amsmath,amssymb,superscriptaddress]{revtex4-1}
\usepackage{graphicx}
\usepackage{dcolumn}
\usepackage{bm}
\usepackage{amsmath}
\usepackage{hyperref} 
\usepackage{soul} 
\DeclareGraphicsExtensions{.pdf,.eps,.png,.jpg,.mps}
\usepackage{mathrsfs}
\usepackage{soul} 
\usepackage[utf8]{inputenc} 
\usepackage{hyperref} 
\usepackage{xcolor} 
\usepackage{eucal}
\usepackage{setspace}
\usepackage{pifont}

\usepackage{flushend}

\usepackage{booktabs}

\begin{document}

\title{Magnetic-thermodynamic phase transition in strained phosphorous-doped graphene}

\author{Natalia Cortés} 
\email{natalia.cortesm@usm.cl}
\affiliation{Instituto de Alta Investigación, Universidad de Tarapacá, Casilla 7D, Arica, Chile}

\author{J. Hernández-Tecorralco}
\affiliation{Instituto de Física, Universidad Nacional Autónoma de México, Apartado Postal 20-364, Ciudad de México C.P. 01000, México.}

\author{L. Meza-Montes}
\affiliation{Instituto de Física, Benemérita Universidad Autónoma de Puebla, Apartado Postal J-48, 72570, Puebla, Puebla, México
}

\author{R. de Coss}
\affiliation{Departamento de Física Aplicada, Centro de Investigación y de Estudios Avanzados del IPN, Apartado Postal 73, Cordemex, 97310, Mérida, Yucatán, México.}
\affiliation{Centro Mesoamericano de Física Teórica, Universidad Autónoma de Chiapas, 29050 Chiapas, México.}

 \author{Patricio Vargas}
\affiliation{Departamento de Física, Universidad Técnica Federico Santa María, 2390123 Valparaíso, Chile}

\date{\today}


\begin{abstract}

We explore quantum-thermodynamic effects in a phosphorous (P)-doped graphene monolayer subjected to biaxial tensile strain. Introducing substitutional P atoms in the graphene lattice generates a tunable spin magnetic moment controlled by the strain control parameter $\varepsilon$. This leads to a magnetic quantum phase transition (MQPT) at zero temperature modulated by $\varepsilon$. The system transitions from a magnetic phase, characterized by an out-of-plane $sp^3$ type hybridization of the P-carbon (P-C) bonds, to a non-magnetic phase when these bonds switch to in-plane $sp^2$ hybridization.
Employing a Fermi-Dirac statistical model, we calculate key thermodynamic quantities as the electronic entropy $S_e$ and electronic specific heat $C_e$. At finite temperatures, we find the MQPT is reflected in both $S_e$ and $C_e$, which display a distinctive $\Lambda$-shaped profile as a function of $\varepsilon$. These thermodynamic quantities sharply increase up to $\varepsilon = 5\% $ in the magnetic regime, followed by a sudden drop at $\varepsilon = 5.5\% $, transitioning to a linear dependence on $\varepsilon$ in the nonmagnetic regime. Notably, $S_e$ and $C_e$ capture the MQPT behavior for low and moderate temperature ranges, providing insights into the accessible electronic states in P-doped graphene. This controllable magnetic-to-nonmagnetic switch offers potential applications in electronic nanodevices operating at finite temperatures.
\end{abstract} 
 
\maketitle
\date{Today}

\section{Introduction}

The transformation of one state of matter into another one driven by temperature $T$ is typically characterized by a phase transition, where both states of matter (or phases) are separated by a boundary and touching at a critical $T$ value. When a phase transition occurs in a magnetic material, the magnetic moment of the electrons plays the major role as $T$ varies, i.e., the orientation of the electron spins changes due to thermal fluctuations induced by $T$. Depending on the magnetic properties of the material (iron, for example), one can observe a magnetic phase transition as a function of temperature, going from a magnetic state into a nonmagnetic state at a critical $T$ value  \cite{PATHRIA1996305}.

At the quantum level, the nature of phase transitions can be different as they manifest in a quantum critical region, where quantum and thermal fluctuations are equally important \cite{sachdev1999quantum}. This type of phase transition starts at absolute zero ($T=0$) and can continue as $T$ increases to some small finite value, as seen in a ferromagnetic phase diagram controlled by an applied magnetic field, for example \cite{sachdev1999quantum}. At $T=0$ the state is denominated quantum phase transition (QPT) \cite{sachdev1999quantum}, and the main feature of a QPT is the so-called quantum critical point (QCP), where thermal fluctuations are suppressed and quantum fluctuations are predominant in the system. The quantum fluctuations are driven by a nonthermal control parameter such as an applied magnetic field, the amount of charge carriers, pressure, or strain, among others \cite{sachdev1999quantum,klanjvsek2014critical}. 

Graphene is a versatile material where magnetic, topological, or quantum phase transitions can appear, which are ruled by diverse control parameters   \cite{vancso2017magnetic,parker2021strain,palma2024entropy,huang2024strain}. Pristine graphene is a nonmagnetic material, and some ways to induce magnetism in it are through defects \cite{yazyev2007defect}, creating samples with defined edges \cite{magda2014room}, or by adding foreign atoms to their lattice \cite{hernandez2022understanding}. It has been shown that phosphorous (P) atoms are good candidates to produce magnetism in graphene by substitutional doping \cite{lin2019p,langer2020tailoring}. Graphene also allows the application of strain at finite temperatures, where its two-dimensional (2D) hexagonal structure possesses exceptional mechanical properties, allowing large deformations without breaking \cite{lee2008measurement}. Uniaxial and biaxial strain are experimentally accessible techniques that can be applied to graphene systems by lattice deformations of its carbon (C) atoms. Strain can be useful as a tool for tuning graphene electronic properties \cite{pereira2009tight}, can serve as a way to assist self-assembly of adsorbed atoms on the graphene lattice \cite{si2016strain}, and can induce a QPT in magic-angle twisted bilayer graphene \cite{parker2021strain}.  

Density functional theory (DFT) studies show that at $T=0$, a P-impurity atom opens a gap in bulk graphene and induces a narrow band at the Fermi level ($E_F$), generating magnetism through a large spin-polarized state with spin splitting of $267$ meV \cite{hernandez2020effects}. This magnetic state is associated with P-C out-of-plane bonds showing $sp^3$-like hybridization in graphene real-space lattice. As tensile strain is applied on P-doped graphene, the $sp^3$ electronic configuration remains up to a certain critical strain value, then the hybridization for the atoms changes to $sp^2$ as strain increases, recovering the flat hexagonal lattice with combined P-C bonds in the constructed supercell. In this process, a magnetic quantum phase transition (MQPT) occurs from a magnetic state to a nonmagnetic state driven by the strain control parameter $\varepsilon$  \cite{hernandez2020effects}.  


One fundamental question may emerge from the abovementioned processes: What will happen with the predicted MQPT in P-doped graphene when temperature is turned on?  
To answer this question, we should know how the entropy can influence the MQPT at $T$ above zero. 
We can access this thermodynamic quantity for electrons through the electronic entropy, $S_e$, and then obtain the electronic specific heat $C_e$, both by employing Fermi-Dirac statistics. These two thermodynamic-electronic quantities are directly linked to each other by temperature, so that they only play a role at finite $T$. 
Experimental measurements of $S_e$ have allowed the acquisition of fundamental information about the accessible electronic states of different systems, such as quantum dots \cite{hartman2018direct} and magic angle twisted bilayer graphene \cite{rozen2021entropic,saito2021isospin}. It was found that in graphene, doping can induce changes in $C_e$ as a function of $T$ \cite{mousavi2013electronic}, and edge states in zigzag graphene nanoribbons can improve $C_e$ at low $T$ \cite{yi2007low}. Only just a few years ago, it was possible to measure $C_e$ in graphene monolayer by using ultrasensitive calorimetric techniques \cite{aamir2021ultrasensitive}. 

In this work, we theoretically predict both thermodynamic quantities $S_e$ and $C_e$ for strained P-doped graphene at finite temperatures. 
We find that $S_e$ and $C_e$ are three orders of magnitude larger than strained pristine graphene. 
We show $S_e$ and $C_e$ reflect the MQPT with a phase transition of a $\Lambda$-type line shape as a function of $\varepsilon$.
We can observe the two characteristic magnetic ($ 0\% \leq \varepsilon \leq 5\% $) and nonmagnetic ($ 5.5\% \leq \varepsilon \leq 10\% $) regimes are still present in the phase diagram at finite $T$ as compared to $T=0$ phase diagram.
We evaluate three different orders of magnitude for $T$, and find the $\Lambda$-type line shape in $S_e$ and $C_e$ is preserved for all $T$.
Interestingly, thermal fluctuations present in $S_e$ and $C_e$ do not destroy the quantum critical region found at $T=0$ phase diagram, instead it is preserved within the same        $\varepsilon$ values for $S_e$ and $C_e$ at finite $T$.

\section{DFT-Thermodynamic Model}

To obtain the electronic and magnetic properties of strained pristine graphene and strained P-doped graphene (the latter labeled as P-graphene), we performed DFT calculations using a plane-wave and pseudopotential method, as implemented in the QUANTUM-ESPRESSO code \cite{giannozzi2009quantum,giannozzi2017advanced}. Valence electrons were represented by plane waves with a kinetic energy cutoff of 55 Ry (320 Ry for the charge density), while core electrons were replaced by ultrasoft pseudopotentials \cite{dal2014pseudopotentials}. We employ the generalized gradient approximation with Perdew–Burke–Enzerhof (PBE) parametrization \cite{perdew1996generalized} for the exchange-correlation functional. A vacuum spacing of 15 \AA\ was used to avoid interaction between periodic images along the $z$-direction. For all cases, atomic positions were relaxed until the internal forces were below 0.01 eV/\AA. Brillouin zone integrations were carried out with a $21\times21$ $k$-point grid \cite{monkhorst1976special} using a Methfessel-Paxton scheme smearing \cite{methfessel1989high} with a width of 0.005 Ry for the constructed unit cells for strained pristine graphene and strained P-graphene.
\begin{figure}[!h]
\centering
\includegraphics[width=\linewidth]{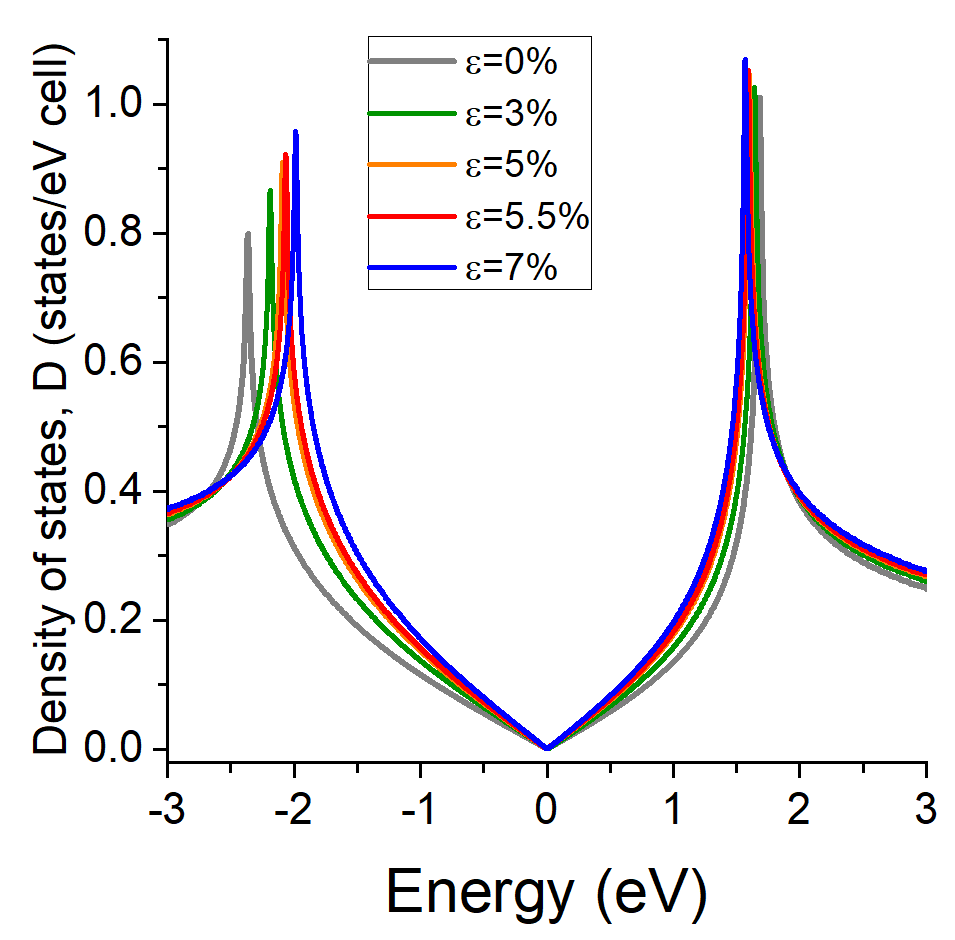}
\caption{DOS for strained pristine graphene where $\varepsilon=0$ represents the unstrained graphene monolayer. The unit cell for pristine graphene has two C atoms. All Fermi levels are set to zero energy.}\label{fig1a} 
\end{figure}
\begin{figure}[!h]
\centering
\includegraphics[width=\linewidth]{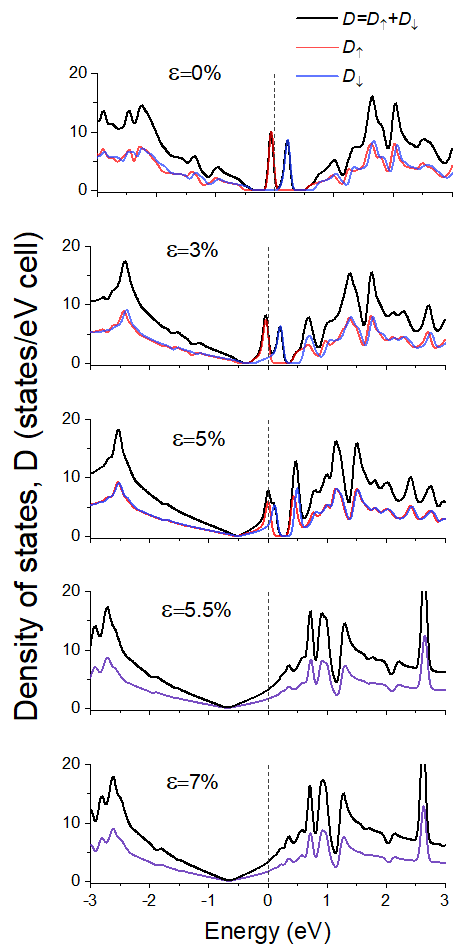}
\caption{DOS for strained P-doped graphene. The unit cell for strained P-graphene has 49 C atoms and 1 P atom. The red color lines represent the majority, and blue color lines represent minority spin contributions in all cases. All Fermi levels are set to zero energy.}\label{fig1b} 
\end{figure}

We simulated P substitutional impurities by replacing one C atom from a graphene layer considering a $5 \times 5$ graphene supercell. Our model consists of 49 C atoms and one P atom, corresponding to $2.0\%$ of impurities concentration of P atoms \cite{hernandez2020effects}. Biaxial tensile strain modulated by the control parameter $\varepsilon$ is applied on the systems by increasing the lattice constant as $a= (1 + \varepsilon)a_0$, where $a_0$ is the unstrained graphene lattice constant, and $\varepsilon$ takes values from 0\% to 10\%. Within these DFT calculations at zero temperature, the electronic density of states (DOS) is obtained for strained pristine graphene and strained P-graphene. The DOS, $D$, we use throughout the paper is the sum of spin up ($\uparrow$) and spin down ($\downarrow$), majority and minority components respectively, $D(E,\varepsilon)=D_{\uparrow}+D_{\downarrow}$. The density of states depends on both the electronic state with energy eigenvalue $E$, and the control parameter $\varepsilon$ applied on either strained graphene system. 

Figure \ref{fig1a} shows the DOS for strained pristine graphene with $\varepsilon$ ranging from $0\%$ to $7\%$. At zero energy [charge neutrality point (CNP)], $D=0$ for each $\varepsilon$ value, then $D$ linearly increases with different slopes around CNP (up to $D\approx$ 0.1 states/eV\ cell). The two van Hove singularities around CNP are nonsymmetric as we use more than one single orbital for C atoms in our DFT calculations.
When a substitutional P impurity atom is added to the monolayer graphene, the behavior of the DOS drastically changes. In Fig.\ \ref{fig1b}, we show $D$ for strained P-graphene, including contributions of $D_{\uparrow}$ and $D_{\downarrow}$ with red and blue lines, respectively. At $\varepsilon=0\%$ (top panel), two large spin-splitting peaks appear around $E_F$ (within the energy range $-0.5 \leq E \leq 0.5$ eV). Each peak corresponding to one type of spin density has contributions of the P-impurity and C atoms of graphene, generating a maximum spin magnetic moment ($M_S$) \cite{GS}. As strain increases, the spin splitting between both spin densities reduces until they become identical at a critical value $\varepsilon=5.5\%$, indicating the change from a magnetic ($D_{\uparrow} \neq D_{\downarrow}$) to a nonmagnetic state ($D_{\uparrow}=D_{\downarrow}$). When $\varepsilon \geq 5.5\%$, the C atoms of the graphene monolayer make room for the P impurity in the hexagonal plane, and the two peaks vanish and merge in the DOS. These latter types of DOS are responsible for the nonmagnetic regime . We will discuss these transitions in the next sections.


These previous DOS calculations at $T=0$ do not provide information about $S_e$ and $C_e$, but we can obtain them through a Fermi-Dirac statistical model as follows. From the constructed graphene supercell, we have that the total number of electrons is $N=201$, in which $N$ must be preserved for each system regardless of the $T$ value. First, we can obtain the chemical potential $\mu(T)$ as a function of $T$ for each $\varepsilon$ value by inversion of
\begin{equation}\label{numelectrons}
N=201=\int_{E_l}^{E_h}D(E,\varepsilon)n_\text{F}(E,T,\mu)dE,
\end{equation}
where $E_{l(h)}$ is the lowest (highest) electronic energy eigenvalue of the considered graphene system, $n{_\text{F}}(E,T,\mu)=1/[e^{\beta(E-\mu(T))}+1]$ is the Fermi-Dirac function distribution with $\beta=1/k_{\text{B}}T$, and $k_{\text{B}}$ is the Boltzmann constant. All the electronic DOS presented in Fig.\ \ref{fig1a} and Fig.\ \ref{fig1b} can be used to obtain $S_e$ and $C_e$ at finite temperature $T>0$. We calculate $S_e$ as  
\begin{equation}\label{entropy}
S_{e}(\varepsilon,T)=-k_{\text{B}} \int_{E_l}^{E_h}D(E,\varepsilon)\mathcal{F}(n_\text{F})dE,
\end{equation}
where
\begin{equation}\label{funcionnumero}
    \mathcal{F}(E,T,\mu)=n_\text{F}\ln n_\text{F}+(1-n_\text{F})\ln(1-n_\text{F}), 
\end{equation}
is approximated by a Lorentzian-like function 
\begin{equation}\label{flore}
    L(E,T,\mu)=\frac{1.4}{e^{(|E-\mu(T)|/2k_{\text{B}}T)^{3/2}}+1},
\end{equation}
where its width is $T$ dependent with full width at half maximum of $\simeq 4k_B T$. By considering low and high $T$, we obtain excellent agreement between Eqs.\ \ref{funcionnumero} and \ref{flore} as $-\mathcal{F}(E,T,\mu)\cong L(E,T,\mu)$. The $L$ function in Eq.\ \ref{flore} plays the role of a filter around $E_F$ for each DOS, as it captures states of the DOS of width $\simeq 4k_B T$. This approximation can be applied to other 2D materials because $L$ is a generic function depending on the system's energy eigenvalues $E$, $\mu$ and $T$, as demonstrated for different quantum structures \cite{cortes2021gate,cortes2022proximity}. Therefore, Eq.\ \ref{entropy} transforms as
\begin{equation}\label{aproxentropy}
S_{e}(\varepsilon,T)\cong k_{\text{B}} \int_{E_l}^{E_h}D(E,\varepsilon)L(E,T,\mu)dE.
\end{equation} 
%

Through Eq.\ \ref{aproxentropy}, we calculate the electronic specific heat $C_{e}$ as
\begin{equation}\label{specheat}
   C_{e}(\varepsilon,T) = T\frac{dS_{e}}{dT}.
\end{equation}
Equations \ref{aproxentropy} and \ref{specheat} can be resolved either as a function of  $\varepsilon$ or $T$. When as a function of $\varepsilon$, we will get information about the accessible electronic states and $\Lambda$-type phase transition for P-graphene at different temperatures.

\section{Finite-temperature results}

\begin{figure*}[!ht]
\centering
\includegraphics[width=\textwidth]{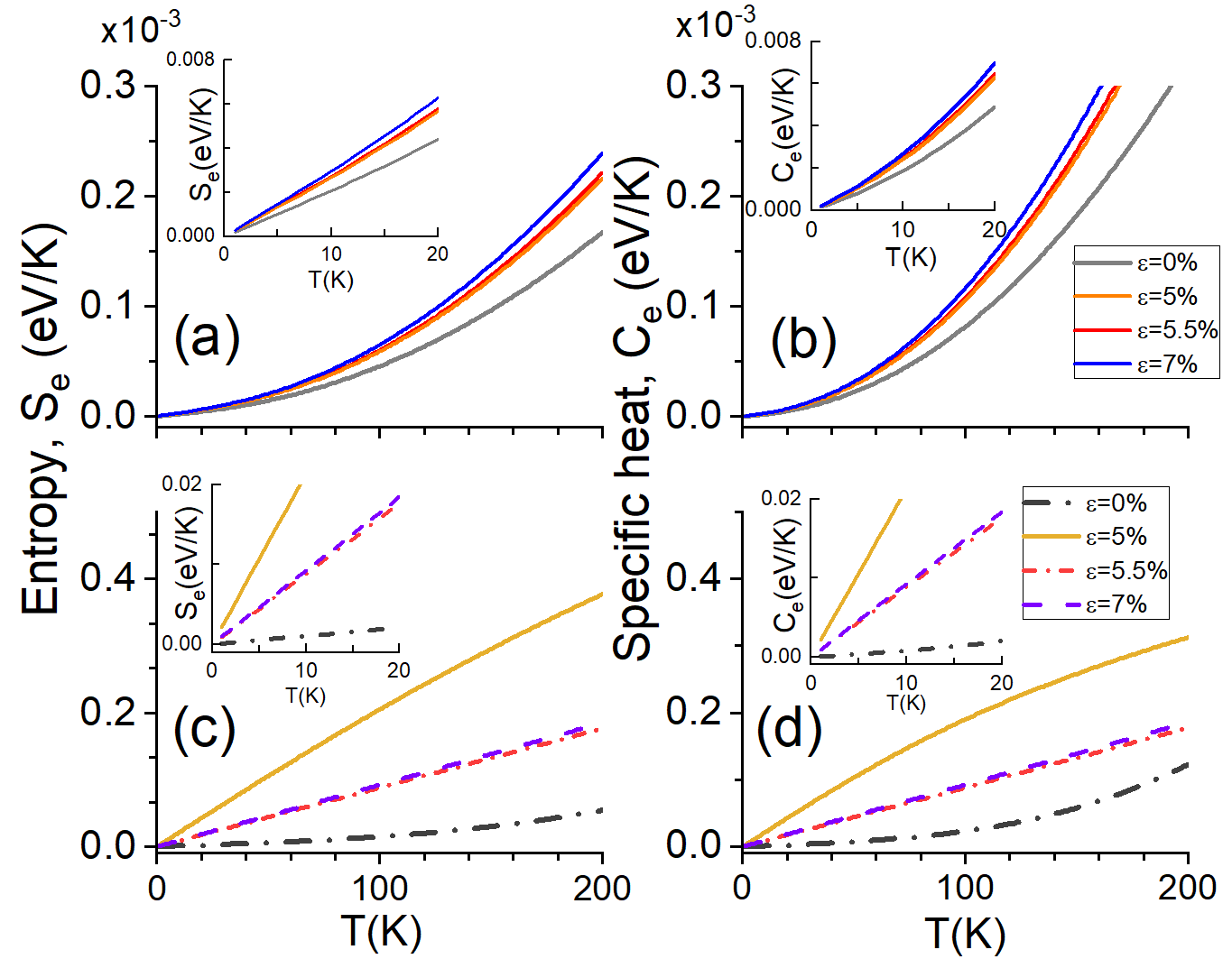}
\caption{(a) and (c) Electronic entropy ($S_e$) calculated through Eq.\ \ref{aproxentropy}; (b) and (d) electronic specific heat ($C_e$) from Eq.\ \ref{specheat}, both per supercell. All quantities as a function of temperature $T$ for different values of $\varepsilon$ as indicated. Top panels (a) and (b): strained pristine graphene, bottom panels (c) and (d): strained P-doped graphene. Insets show a zoom for each quantity with $T$ going from 1 K to 20 K. Vertical scales in (a) and (b) and their respective insets times $10^{-3}$. Notice $\varepsilon=0\%$ corresponds to unstrained graphene in each plot.}\label{fig2} 
\end{figure*}

Figure \ref{fig2} shows $S_e$ in left panels, and $C_e$ in right panels for strained pristine graphene (top panels) and strained P-graphene (bottom panels).
We emphasize that the thermodynamic quantities are calculated per unit cell in each case \cite{units}. When pristine graphene is biaxially strained, $S_e$ and $C_e$ show similar behavior as seen in panels (a) and (b). Both quantities monotonically increase as $T$ increases, but $C_e$ increases faster than $S_e$, and their lowest values occur for the unstrained case $\varepsilon=0$, while the maxima are for $\varepsilon=7\%$. The inset in Fig.\ \ref{fig2}(a) shows that $S_e$ is a linear function of $T$ up to $T\simeq 20$ K, and  $C_e$ is linear with $T$ up to $T\simeq8$ K as shown in the inset of Fig.\ \ref{fig2}(b). These monotonically thermodynamic responses for strained pristine graphene do not show significant variations as $\varepsilon$ changes and $T$ increases.   

However, $S_e$ and $C_e$ substantially change their behavior for strained P-graphene, as shown in Fig.\ \ref{fig2}(c) and (d), respectively. Both quantities are three orders of magnitude larger than for strained pristine graphene due to the contribution of the P-impurity atom states. For P-graphene, the $L$ function in Eq. \ref{flore} captures more available states of each DOS as $T$ increases, as one can infer from Fig.\ \ref{fig1b}. These captured states are mainly due to the spin-polarized peaks around $E_F$ for each strain value. The highest $S_e$ and $C_e$ magnitudes occur for $\varepsilon=5\%$ instead of $\varepsilon=7\%$ as in the pristine case. For $\varepsilon=5\%$, the $L$ function captures a maximum around $E_F$, see the mid panel in Fig.\ \ref{fig1b}, indicating the highest quantity of available electronic states occur at $\varepsilon=5\%$. When $\varepsilon>5\%$, $S_e$ and $C_e$ linearly increase as $T$ increases. For these strain values ($\varepsilon>5\%$), one can see from the DOS in Fig.\ \ref{fig1b} that the peak states are no longer distinguishable near $E_F$, therefore the $L$ function captures less available states and $S_e$ and $C_e$ are lower than for $\varepsilon=5\%$. At $T\leq 20$ K, $S_e$ and $C_e$ show high similarity, linearly increasing with $T$ as shown in insets of Fig.\ \ref{fig2}(c) and (d). The non monotonically behavior for $S_e$ and $C_e$ as a function of $T$, and the sudden jump at $\varepsilon=5\%$ for all $T$, indicates that a phase transition can be taking place in strained P-graphene at finite $T$.

\begin{figure*}[!ht]
\centering
\includegraphics[width=\textwidth]{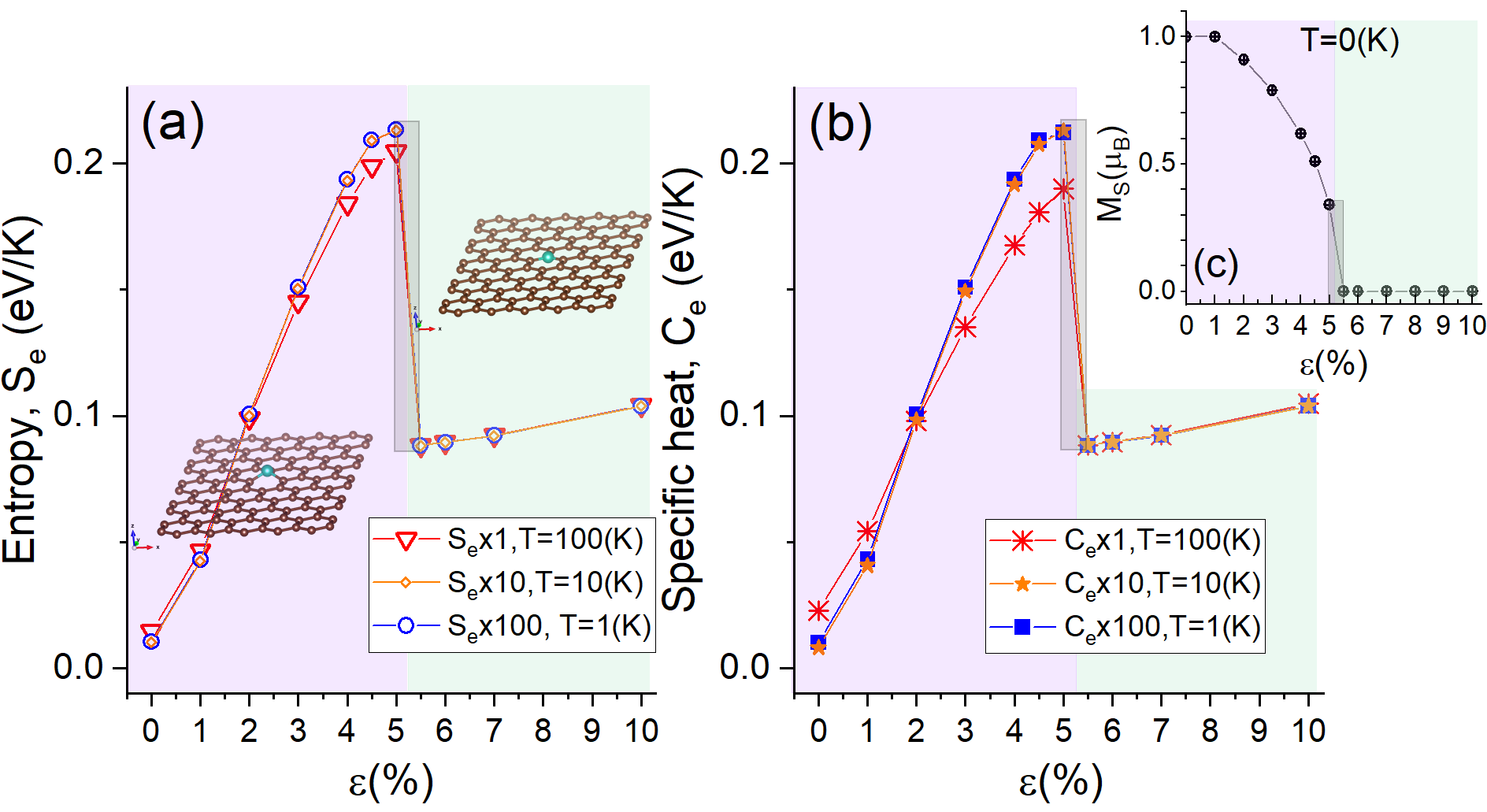}
\caption{(a) Electronic entropy $S_e$, (b) electronic specific heat $C_e$, (c) spin magnetic moment ($M_\text{S}$) at $T=0$ K. All quantities calculated per supercell and as a function of the control parameter $\varepsilon$ for strained P-graphene. $S_e$ and $C_e$ include results for three temperatures, 1, 10, and 100 K, where we have amplified each quantity by 100, 10, and 1, respectively, to superimpose the data. Violet zone for each plot in the range $0\% \leq \varepsilon < 5.5\%$, corresponding to the magnetic phase, and green region $5\% < \varepsilon \leq 10\%$ in the nonmagnetic phase. Shading gray rectangles delimit the critical transition region ($5\% \leq \varepsilon \leq 5.5\%$). In (a) electronic configuration schematics for $sp^3$ (magnetic phase) and $sp^2$ (nonmagnetic phase), where brown spheres represent C atoms in graphene and the green sphere around its center the P atom.}\label{fig3}
\end{figure*}

To get insight into that peculiar behavior for strained P-graphene, we present in Fig.\ \ref{fig3}, $S_e$ in panel (a), and $C_e$ in panel (b), both as a function of $\varepsilon$ and three different $T$ values, $T=1$ K, $T=10$ K and $T=100$ K. We choose the curve for $T=100$ K [red triangles in panel (a) and red asterisk symbols in panel (b)] for the following description; however, the same applies as $T$ decreases up to $T=1$ K (with the exception of their numerical magnitudes). We also include results for the spin magnetic moment, $M_\text{S}$, in panel (c), to highlight the MQPT at $T=0$ K. All these results are particularly interesting due to several factors.

At finite temperatures, both $S_e$ and $C_e$ exhibit a strikingly similar behavior. Their respective line shapes increase sharply with rising $\varepsilon$, reaching a peak value of $\approx 0.2$ eV/K at $\varepsilon = 5\%$. Beyond this point, both quantities suddenly drop at $\varepsilon = 5.5\%$, with magnitude of $\approx0.1$ eV/K. For $\varepsilon \geq 5.5\% $, $S_e$ and $C_e$ gradually increase with $\varepsilon$, displaying minor variations in their linear profiles as $T$ rises. This overall behavior, characterized by a $\Lambda$-like line shape that remains largely consistent with changing $T$ \cite{superphase}, suggests the presence of two distinct regimes when in comparison with the MQPT observed at $T = 0$, as depicted in Fig.\ \ref{fig3}, panel (c).

In Fig.\ \ref{fig3}(c), we show the MQPT with $M_\text{S}$ as a function of $\varepsilon$ at $T=0$ \cite{hernandez2020effects}, where we can identify two regimes, the first one is the magnetic phase, in the range $0\%\leq\varepsilon \leq 5.5\%$ (violet color area). In the second regime in the range $5.5\%\leq\varepsilon\leq10\%$ (green color area), a nonmagnetic phase ($M_\text{S}=0$) is seen. This MQPT is closely related to the electronic configuration of the electron states that contribute to $S_e$ and $C_e$. When $\varepsilon < 5.5\%$, the substitutional P impurity atom is positioned above the graphene plane because the P atom does not fit into the unstrained ($\varepsilon=0$) graphene flat hexagonal lattice. However, when $\varepsilon \geq 5.5\%$, the P impurity aligns within the same plane as graphene, transitioning from an $sp^3$-like to an $sp^2$ electronic configuration, see atomic schemes in Fig.\ \ref{fig3}(a).
This transition causes the spin-polarized state for the P atom to go to zero ($M_\text{S}=0$) as $\varepsilon \geq 5.5\%$, leading the system from a magnetic phase into a nonmagnetic one, and the MQPT is manifested in strained P-graphene at $T=0$.



As Fig.\ \ref{fig3} panels (a) and (b) show, $S_e$ and $C_e$ increase for the same strain values when the magnetism goes down at $T=0$, see violet regions in all panels of Fig.\ \ref{fig3}. In this regime, the strained P-graphene system increases the quantity of available electronic states as long as the P-atom induces magnetism up to $\varepsilon=5\%$. Then $S_e$ and $C_e$ abruptly drop at $\varepsilon=5.5\%$. Since $S_e$ and $C_e$ involve electronic states within a small energy range of width $\simeq 4k_B T$, and very close to $E_F$, the phenomenon of the transition between $\varepsilon=5\%$ and $\varepsilon=5.5\%$ is evident in the DOS shown in Fig.\ \ref{fig1b}. The DOS around $E_F$ for $\varepsilon=5\%$ is completely different compared to the DOS for $\varepsilon=5.5\%$. The first one shows a peak, and the latter a small curvature around $E_F$; the DOS significantly decreases between these two strain values. This explains the abrupt drop for $S_e$ and $C_e$ at $\varepsilon=5.5\%$. Right at the drop of $S_e$ and $C_e$, both thermodynamic quantities remain proportional to $\varepsilon$ with a straight line shape, and the magnetism vanishes in this phase.


Notably, $S_e$ and $C_e$ reveal a critical region for $\varepsilon$ in the range $5\%\leq\varepsilon\leq 5.5\%$ (see shading gray rectangle in each plot), just when the system transitions from a magnetic phase to a nonmagnetic one (or vice-versa) even at temperatures higher than zero, where $S_e$ and $C_e$ have magnitudes of $\approx 0.1$ eV/K at $T=100$ K. This critical region can tell us that there is a mixing of quantum and thermal fluctuations competing to lower the electronic entropy and specific heat to reach a stable state for the system.         
In other words, the thermodynamic quantities $S_e$ and $C_e$ for $T \neq 0$ may indicate quantum criticality within this region \cite{sachdev1999quantum}. We highlight that the thermodynamic quantities reported here are for temperatures up to $T = 200$ K. Although extending the analysis to higher temperatures is feasible, our focus is based on electronic specific heat experiments \cite{aamir2021ultrasensitive}, which are done within this $T$ range, capturing the behavior for moderate temperatures.

\section{Conclusions}

Strained phosphorous-doped graphene exhibits emergent magnetism as long as a $sp^3$-like hybridization of the P-C bonds takes place in the system. The applied strain control parameter $\varepsilon$, plays a critical role in modulating the $sp^3$-like electronic configuration at absolute zero ($T=0$), as well as influencing thermodynamic quantities such as the electronic entropy $S_e$, and electronic specific heat $C_e$ at finite temperatures $T \neq 0$. 

At $T=0$, the system undergoes a magnetic quantum phase transition (MQPT) driven by $\varepsilon$, shifting from a magnetic state characterized by an $sp^3$ hybridization ($0\% \leq \varepsilon \leq 5\%$) to a nonmagnetic state with $sp^2$ hybridization ($5.5\% \leq \varepsilon \leq 10\%$) where the phosphorus atom becomes coplanar with the graphene sheet. At non-zero temperatures, the behavior for $S_e$ and $C_e$ when as a function of $\varepsilon$ reflects the MQPT, displaying a distinctive $\Lambda$-lineshape response that persists for temperatures around $100$ K.

The quantities $S_e$ and $C_e$ are particularly effective in distinguishing between the magnetic and nonmagnetic regimes at finite temperatures, corresponding to the same strain values where the MQPT is observed at $T=0$. The transition between these two regimes defines a critical strain region, approximately in the range $5\% \leq \varepsilon \leq 5.5\%$, where a competition between quantum and thermal fluctuations emerges. The $\Lambda$-type phase transition identified here is crucial for understanding the accessible states near the Fermi level. This behavior could be experimentally probed via the thermodynamic responses of $S_e$ and $C_e$, providing insights into magnetic quantum phase transitions occurring for temperatures above zero.



\section{Acknowledgments}

N.C. acknowledges support from ANID Iniciación en Investigación Fondecyt Grant No. 11221088 and DGII-UTA, and the hospitality of Universidad Federico Santa María, Valparaíso, Chile. J.H.-T. acknowledges a postdoctoral fellowship from CONAHCyT-México. The authors thankfully acknowledge the computer resources, technical expertise, 
and support provided by the Laboratorio Nacional de Supercómputo del Sureste de México(LNS), a member of the CONACYT national laboratories, with project No. 202303063N.
One of the authors (R. de Coss) is grateful for the hospitality of the Mesoamerican Centre for Theoretical Physics (MCTP), where part of this work was developed during a research visit.
P.V. wishes to thank the Fondecyt grant project No. 1240582.

\bibliography{bib}

\begin{thebibliography}{34}
\bibitem [{GS()}]{GS}%
  \BibitemOpen
  \href@noop {} {}\bibinfo {howpublished} {Note that the ground state occurs at
  $\varepsilon=0\%$ with $M_\text{S}=1$ $\mu_B$ per supercell.}\BibitemShut
  {Stop}%
\bibitem [{uni()}]{units}%
  \BibitemOpen
  \href@noop {} {}\bibinfo {howpublished} {Since entropy and specific heat are
  extensive properties, they can be easily normalized using appropriate
  factors, such as per atom or in J/(mol K). In this study, we present the
  entropy and specific heat values on a per-unit-cell basis, reported in units
  of eV/K.}\BibitemShut {Stop}%
\end{thebibliography}

\end{document}